\documentstyle[epsfig,prc,aps,12pt]{revtex}

\begin {document}
\tighten
\title{Alpha particle production by molecular single-particle 
effect \\ in reactions of $^{9}$Be just above the Coulomb barrier}
\author{A. Diaz-Torres$^{a,}$$\footnote{Corresponding author.\\ 
{\it E-mail address}: A.Diaz-Torres@surrey.ac.uk\\
{\it Fax number}: +44 (0)1483 686781}$, 
I.J. Thompson$^{a}$ and W. Scheid$^{b}$}
\address{$^{a}$Physics Department, University of Surrey, Guildford GU2
7XH, United Kingdom\\
$^{b}$Institut f\"ur Theoretische Physik der Justus--Liebig--Universit\"at, 
D--35392 Giessen, Germany}
\date{\today}
\maketitle

\begin{abstract}
The $\alpha $-particle production in the
dissociation of $^{9}$Be on $^{209}$Bi and $^{64}$Zn at energies just above 
the Coulomb barrier
is studied within the two-center shell model approach. The dissociation of  
$^{9}$Be on $^{209}$Bi is caused by a molecular single-particle effect 
(Landau-Zener mechanism) before the nuclei reach the Coulomb barrier. 
Molecular single-particle effects do not occur
at that stage of the collision for $^{9}$Be+$^{64}$Zn, and
this explains the absence of fusion suppression observed for this system.
The polarisation of the energy level of the last neutron of $^{9}$Be 
and, therefore the existence of avoided crossings 
with that level, depends on the structure of the target.
\end{abstract}



$\emph {Introduction:}$
The study of breakup processes of weakly bound or halo
nuclei at energies around the Coulomb barrier (\cite{Kolata1} and references
therein) is a very lively topic due to the increasing interest in nuclear
reactions with radioactive beams. The effect of breakup on fusion and
scattering has also been extensively investigated in recent years both
theoretically \cite{TH4,Yabana,hagino,Alexis,Alexis1} and
experimentally \cite{Signorini1,Lubian}.

In a recent paper \cite{Alexis}, we have studied the 
reaction $^{9}$Be+$^{209}$Bi at energies above and near the Coulomb barrier 
in the adiabatic
two-center shell model (TCSM) \cite{Maruhn} approach. The effect of
dissociation of $^{9}$Be$\rightarrow $n$+2\alpha $ on complete fusion and
scattering processes was particularly investigated. Results showed that the
dissociation of $^{9}$Be could be due to a molecular single-particle effect
at two very close avoided crossings between the state $j_{z}=3/2$ of the
last neutron of $^{9}$Be and two unoccupied states of $^{209}$Bi with the
same projection of the single-particle total angular momentum $j_{z}$ on the
internuclear axis, before the nuclei reach the Coulomb
barrier (Fig. 1). The dissociation of $^{9}$Be is caused by two very
competitive (simultaneous) transitions of the last neutron of $^{9}$Be at
these avoided crossings (1-2) induced by the radial motion of the nuclei 
\cite{Greiner}, namely into the continuum (1) and to $^{209}$Bi (2). These
transitions lead to the formation of the unbound nucleus $^{8}$Be which may
decay immediately into two $\alpha $-particles. The $\alpha $-particles 
originating from the decay of $^{8}$Be have
been experimentally observed in this reaction \cite{Signorini}. 

The aim of this letter is to compare the $\alpha $-particle production in the
dissociation of $^{9}$Be on $^{209}$Bi and on $^{64}$Zn, 
at energies just above the Coulomb barrier,
within the TCSM \cite{Maruhn} approach. 
The use of adiabatic TCSM basis states is well justifed \cite{Greiner} at 
low energy nuclear reactions (near the Coulomb barrier), where the relative 
motion of 
the nuclei is slow compared with the rearrangement time of the mean field of 
nucleons. The single-particle motion of valence nucleons during the reaction 
is described by molecular orbitals which are polarised by the field of the 
other nucleus. Once a neck in the dinuclear system starts to form and 
molecular single-particle levels develop, 
there can occur transitions \cite{Greiner} (molecular single-particle effects) 
between those levels, induced by the relative motion of the nuclei 
(operators associated with the kinetic energy). 
In particular, transitions at the avoided crossings of molecular levels, 
induced by the radial motion of the nuclei 
(Landau-Zener mechanism \cite{Greiner}), describe the non-adiabaticity of the 
radial motion and grow with increasing radial velocity, the diabatic 
single-particle motion being a limit for large radial 
transition probabilities \cite{Cassing}. In the diabatic limit of the 
single-particle motion \cite{Cassing}, the nucleons do not occupy the lowest 
free single-particle levels as in the adiabatic case, but remain in 
the diabatic levels keeping their quantum numbers during a collective motion 
of the nuclear system. In this letter, we will deal with an intermediate 
situation between the adiabatic and diabatic limits of the single-particle 
motion. 

It would be interesting to calculate the 
cross sections for the $\alpha $-particle production in the approach 
phase of these collisions and to compare them with the experimental data, 
because this study may or may not support
the dissociation of $^{9}$Be by molecular single-particle
effects \cite{Alexis}.
Moreover, we expect that the dissociation of $^{9}$Be and, therefore, 
the existence of a suppression mechanism for fusion, could 
be associated with the evolution of the last neutron of $^{9}$Be 
in the approach phase of the collision before the nuclei reach the Coulomb 
barrier, and this may depend on the structure of the partner nucleus. 
The study of the reaction $^{9}$Be+$^{64}$Zn is interesting because there 
is no evidence for fusion suppression for 
this system \cite{Lubian}, in contrast to the 
reaction $^{9}$Be+$^{209}$Bi \cite{Signorini1}.  

$\emph {Model:}$
The cross section $\sigma$ for the production of $\alpha$-particles 
is defined as 
\begin{equation}
\sigma=\frac{\pi }{2\mu \text{E}_{\text{c.m.}}}\sum_{l=0}^{l_{max}}(2l+1)(2\cdot P(\text{E}_{\text{c.m.}},l)),
\label{Eq1}
\end{equation}
where $l_{max}$ is a maximum orbital angular momentum for the relative
motion of the nuclei which can reach the avoided crossing of molecular 
levels  at the incident
energy E$_{\text{c.m.}}$, $\mu$ denotes the reduced mass for the relative motion of the
nuclei, and $P(\text{E}_{\text{c.m.}},l)$ is the dissociation probability of $^{9}$Be
(production of $^{8}$Be) for the incident energy E$_{\text{c.m.}}$ and orbital angular
momentum $l$.
The factor 2 appears in Eq. (\ref{Eq1}) because two $\alpha $-particles are
produced in the decay of a $^{8}$Be nucleus. 
 The existence of the dissociation channel for $^{9}$Be 
(a neutron transition into the continuum or a neutron transfer) is derived, in 
the present model, from examining the neutron levels calculated with the TCSM.

In the complete fusion process ($0\leq l\leq l_{cr}$), the dissociation
probability at the avoided crossing of molecular adiabatic single-particle 
levels can be expressed as 
\begin{equation}
P=P^{(e)},  \label{LZ4}
\end{equation}
and in the scattering process ($l_{cr}\leq l\leq l_{max}$) as 
\begin{equation}
P=P^{(e)}+(1-P^{(e)})\cdot P^{(lv)},  \label{LZ3}
\end{equation}
since the system passes through the avoided crossing point twice, first
entering $(e)$ and then leaving $(lv)$ the interaction region. Here, we have
made a distinction between $(e)$ and $(lv)$ because the radial velocity of
the nuclei changes due to frictional effects \cite{Alexis}. The first term
in Eq. (\ref{LZ3}) is $P^{(e)}$, instead of $P^{(e)}\cdot (1-P^{(lv)})$,
because of our assumption of an immediate dissociation of $^{8}$Be $\rightarrow
2\alpha $ after the neutron in $^{9}$Be has been removed from the nucleus
entering the interaction region (the neutron does not get bound again with $%
^{8}$Be when the nuclei leave the interaction region). 
The dissociation
probability $P^{(e),(lv)}$ at an isolated avoided crossing of molecular 
levels is calculated with the Landau-Zener formula \cite{Greiner}. 
The radial velocity $v^{0}$, which is the 
only quantity concerning relative motion in the above expression, is 
calculated from the classical equations of motion for the nucleus-nucleus
potential \cite{Alexis} obtained with the TCSM and the Strutinsky method.
Since the radial velocity $v^{0}$ depends on 
the incident energy E$_{\text{c.m.}}$ and the
orbital angular momentum $l$, $P_{LZ}$ depends also 
on E$_{\text{c.m.}}$ and $l$. 
In general, the neutron transition at the avoided crossing can be
calculated semiclassically \cite{Greiner} using a 
time-dependent Schr\"{o}dinger
equation for the neutron wave function expanded in a basis of two diabatic
wave functions related to the diabatic levels $\epsilon _{1},\epsilon _{2}$
and a classical equation for the relative motion of the nuclei. 

$\emph {Results and discussion:}$ For the calculation of the single-particle levels with the TCSM \cite{Maruhn} for the systems $^{9}$Be+$^{209}$Bi and $^{9}$Be+$^{64}$Zn, we use \cite{Alexis} 
(i) the experimental nucleon separation energies \cite{Sn1,Sn2,Sn3,Sn4,Sn5} for
the colliding nuclei and for the compound nucleus to obtain the depths of the
two oscillator potential wells, (ii) the parameters $\kappa$ and $\mu$ of 
the Nilsson model for the spin-orbit interaction \cite{Ring} and (iii) a set 
of universal parameters, e.g., the nuclear-radius 
constant $r_{0} = 1.2249$ fm and 
the oscillator quanta $\hbar \omega_{i0}=41\cdot A_{i}^{-1/3}$ MeV. 
Moreover, we have considered
spherical nuclei with a value of the neck parameter $\varepsilon =0.75$.
With this value of $\varepsilon$, the neck radius and the internuclear
distance at the touching configuration are approximately equal to those
in the dinuclear system formed by the overlap of the two nuclear frozen
densities. 

($^{9}$Be+$^{209}$Bi) 
Fig. 2 (upper part) shows the dissociation probability $P(\text{E}_{\text{c.m.}}%
,l)$ of $^{9}$Be on $^{209}$Bi as a function of the orbital angular 
momentum $l$ ($0\leq l\leq l_{max}$) for two values of the incident 
energy E$_{\text{c.m.}}$, namely $44.1$ MeV
(solid curve, $l_{cr}=8$ and $l_{max}=14$) and $57.5$ MeV (dashed curve, $%
l_{cr}=16$ and $l_{max}=26$). The dissociation
probability $P^{(e),(lv)}$ in Eqs. (\ref{LZ4}-\ref{LZ3}) is calculated as 
the product of 
two isolated Landau-Zener transitions at the avoided crossings 1 and 2 
of Fig.1. 
At these avoided crossings, the applicability \cite{Greiner} of the 
Landau-Zener approach is quite good. The values of $\mid H_{12}^{\prime
}\mid $ and $\frac{d}{dr}(\epsilon _{1}-\epsilon _{2})$ at the avoided
crossings 1-2 are $0.1073$ MeV, $0.0184$ MeVfm$^{-1}$ and $0.1250$ MeV, $%
0.0352$ MeVfm$^{-1}$, respectively.
For a fixed incident energy E$_{\text{c.m.}}$,
the dissociation probability decreases with increasing orbital angular
momentum $l$ because the potential energy increases and the radial velocity $%
v^{0}$ (kinetic energy) decreases at the avoided crossing points. 
For a fixed orbital angular momentum $l$, the radial velocity $v^{0}$ at
the avoided crossing points and, therefore, the dissociation probability
increases with an increasing incident energy.

Fig. 2 (lower part) shows the calculated excitation function for the 
production of $\alpha $-particles (solid curve) for $^{9}$Be+$^{209}$Bi. 
The calculated cross sections correspond to 
twice the cross sections for the production of $^{8}$Be 
by breakup (avoided crossing 1 of Fig. 1) and 
transfer (avoided crossing 2 of Fig. 1). 
Calculated cross sections for the production of $\alpha $-particles 
underestimate the experimental data 
for $^{9}$Be+$^{209}$Bi reported in \cite{Signorini} by a 
factor 3.7-4.7 in the low range of incident 
energies studied, but agree with 
preliminary experimental data for the similar 
system $^{9}$Be+$^{208}$Pb reported in \cite{Fulton}. 
However, the experimental complete fusion cross sections \cite
{Signorini2,Dasgupta} are similar in both reactions, and 
the fusion cross sections agree well
with those values calculated in the TCSM approach \cite{Alexis}. 
The experimental total breakup+transfer 
cross sections for $^{9}$Be+$^{209}$Bi reported in \cite{Signorini} 
were obtained by integrating over angles the angular distribution 
of $^{8}$Be (i.e. $2\alpha$) nuclei, while both breakup and transfer cross 
sections for $^{9}$Be+$^{208}$Pb reported in \cite{Fulton} were separately 
obtained from the $\alpha$-particles with the beam velocity. 
It is important to note both that the total breakup+transfer cross 
sections 
reported in \cite{Signorini} were obtained from a bump of 
$\alpha$-particles in the charged particles spectra for 
$^{9}$Be+$^{209}$Bi observed systematically in three experiments 
\cite{Signorini} and that therefore, there may be some uncertainties 
about these 
cross sections \cite{Signorini3}. 
New experiments focused on this particular problem seem to be necessary to 
clarify the experimental cross
sections for the production of $\alpha $-particles in the studied reaction
and, therefore, the dissociation of $^{9}$Be by molecular single-particle 
effects.

Since the avoided crossings 1-2 of Fig. 1 have similar
features and the neutron transitions at these avoided crossings occur
simultaneously, we expect that cross sections for the neutron transition
into the continuum (usual breakup) are similar to those for the
neutron transition to $^{209}$Bi (transfer). This expectation agrees well with 
preliminary experimental breakup and transfer excitation functions 
obtained for $^{9}$Be+$^{208}$Pb \cite{Fulton}, which show similar features. 
In the present approach, we cannot separately calculate cross sections for the
neutron transition either into the continuum or to $^{209}$Bi in a realistic
way because both processes interfere strongly with each other. 

For incident energies very close to 
the Coulomb barrier, we expect a $\it{knee}$ in the excitation function 
for the production of $\alpha $-particles because the 
radial velocity
(kinetic energy) of the nuclei and, therefore, the dissociation probability
should increase with a decreasing nucleus-nucleus potential \cite{Alexis} 
(e.g., pole-to-pole orientation) due to deformation and orientation 
effects of $^{9}$Be. 
A more general TCSM \cite{Nuhn} would have to be used for arbitrary
orientations of the intrinsic symmetry axes of deformed nuclei as well as a
quantum mechanical description of the relative motion of the nuclei for
incident energies below the Coulomb barrier \cite{Thiel}. 

In the present work, the calculation of the cross sections 
for the production of $\alpha $-particles is 
focused on incident energies above the Coulomb barrier ($V_{B}\approx 40$
MeV \cite{Alexis}) for two reasons, namely we consider spherical nuclei,
which model is not suitable for incident energies very close to and below the
Coulomb barrier, and the dissociation probabilities cannot be calculated
using the Landau-Zener formula (\ref{LZ1}) for subcoulomb trajectories
because a radial velocity of the nuclei is not defined in the classically
forbidden region. The assumption of spherical nuclei is suitable if there is 
only a small change of the nucleus-nucleus potential, due to orientation and 
deformation effects of the nuclei, in comparison with the surplus of incident
energy above the potential.

($^{9}$Be+$^{64}$Zn) 
Fig. 3 shows the neutron level diagram of the TCSM for $^{9}$Be+$^{64}$Zn 
around the radius of the Coulomb barrier (the B arrow). 
The internal arrow (A) indicates the
distance $r_{t}\approx 7.45$ fm corresponding to the touching configuration 
of the nuclei if no neck is formed and the C arrow indicates 
the relative distance $r\approx 9.21$ fm
where a neck between the nuclei starts to form. 
The values of the Coulomb barrier and its position, calculated with 
the TCSM and the Strutinsky method \cite{Alexis}, 
are $V_B \approx 15.5$ MeV and $r_B\approx 9.43$ fm, respectively. 
These values agree well with the barrier parameters determined 
in \cite{Lubian} from 
the experimental fusion excitation function, i.e. 
$V_{B}^{exp} = 16.2$ MeV and $r_{B}^{exp} = 10$ fm. 

From Fig. 3, we can see that no neck between the nuclei is formed before 
the nuclei reach the Coulomb barrier and, therefore, molecular effects 
cannot occur in that phase of the collision. 
Neutron transitions between the state $j_{z}=3/2$ of the
last neutron of $^{9}$Be and unoccupied states 
of $^{64}$Zn (neutron transfer) could occur at 
relative distances well inside the Coulomb barrier, near the 
touching configuration of the nuclei if no neck is formed (the A arrow). 
In this case the nucleus $^{8}$Be may be completely absorbed by 
the nucleus $^{65}$Zn. On the other hand, it could be expected that 
those transition probabilities caused by a Landau-Zener mechanism 
are very small because the radial velocity decreases at 
the touching configuration of the nuclei due to 
strong frictional effects \cite{Alexis}. 
From these results, we do not expect significant 
production of $\alpha$-particles in the approach phase of this collision. 
This also 
explains the absence of fusion suppression observed for this 
system \cite{Lubian}. Comparing Figs. 1 and 3, it is 
observed that the energy level $j_{z}=3/2$ of the last neutron of $^{9}$Be 
polarises in a different way when $^{9}$Be approaches 
either $^{209}$Bi (goes up) or $^{64}$Zn (goes down). This shows that 
the polarisation of the energy level of the last neutron of $^{9}$Be 
and, therefore the existence of avoided crossings with that level, 
depends on the structure of the target.  
 
Fig. 4 shows the complete fusion excitation function (solid curve) 
for $^{9}$Be+$^{64}$Zn calculated 
within the Glas-Mosel model \cite{Mosel} and using the nucleus-nucleus 
potential obtained with the TCSM and the Strutinsky method. 
The calculated complete fusion cross 
sections agree well with the experimental data \cite{Lubian} (full squares).        

$\emph {Summary and conclusions:}$
The $\alpha $-particle production in the
dissociation of $^{9}$Be on $^{209}$Bi and $^{64}$Zn at energies just above 
the Coulomb barrier
has been studied within the two-center shell model approach. 
The dissociation of  
$^{9}$Be on $^{209}$Bi is caused by a molecular single-particle effect 
(Landau-Zener mechanism) at two very close avoided crossings between 
the state of the last
neutron of $^{9}$Be and two unoccupied states of $^{209}$Bi before
the nuclei reach the Coulomb barrier. The dissociation probability of $^{9}$%
Be decreases with decreasing incident energies, while it 
increases with decreasing orbital angular momentum for a fixed incident 
energy. 
Calculated cross sections for the production of $\alpha $%
-particles in $^{9}$Be+$^{209}$Bi underestimate the available 
experimental data, 
but agree with preliminary experimental data for the 
similar system $^{9}$Be+$^{208}$Pb. New
experiments seem to be necessary to clarify the experimental cross sections
for the production of $\alpha $-particles in $^{9}$Be+$^{209}$Bi and,
therefore, the dissociation of $^{9}$Be by molecular single-particle 
effects. 
We expect similar excitation functions for the neutron transition either 
into the continuum (usual breakup) or to $^{209}$Bi (transfer). 
We also expect a $\it{knee}$ in the excitation function 
for the production of $\alpha $-particles in $^{9}$Be+$^{209}$Bi at 
energies very close to the Coulomb barrier. 
We expect little production of $\alpha $-particles 
for $^{9}$Be+$^{64}$Zn. Molecular single-particle effects do not occur 
before these nuclei reach the Coulomb barrier, and this explains the absence 
of fusion suppression observed for this system. 
The complete fusion excitation function for this system, calculated within the 
Glas-Mosel model and using the nucleus-nucleus potential obtained with 
the two-center shell model and the Strutinsky method, agrees well with 
the experimental data. The energy level $j_{z}=3/2$ of the last 
neutron of $^{9}$Be polarises in a different way when $^{9}$Be approaches 
either $^{209}$Bi (goes up) or $^{64}$Zn (goes down). 
The polarisation of the energy level of the last neutron of $^{9}$Be 
and, therefore the existence of avoided crossings 
with that level, depends on the structure of the target.
   
{\bf Acknowledgments}
 
We thank Prof. Cosimo Signorini for fruitful discussions and comments. 
UK support from the EPSRC grant GR/M/82141 is acknowledged.

\newpage
\begin{figure}
\begin{center}
\epsfig{file=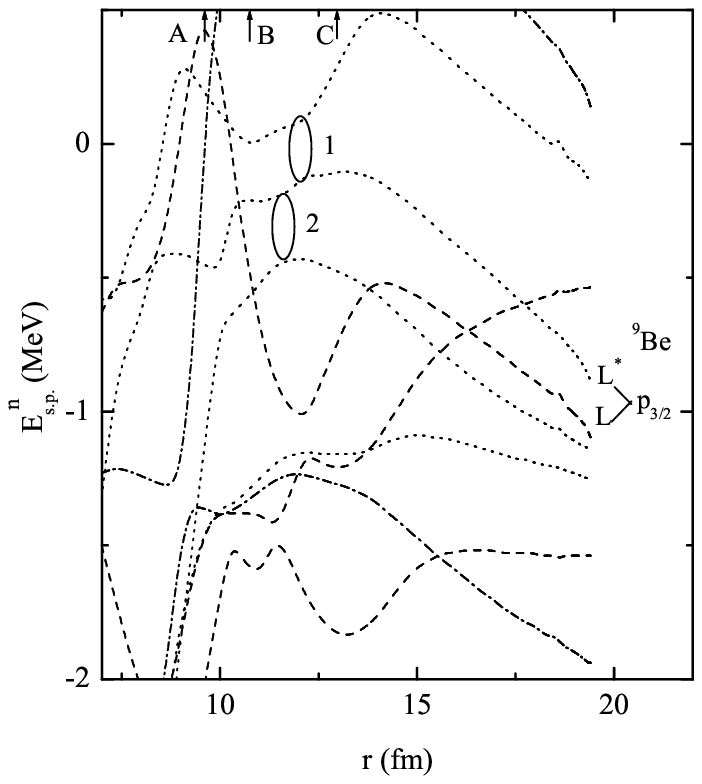,width=0.75\textwidth}
\end{center}
\caption{Neutron levels for $^{9}$Be + $^{209}$Bi $\rightarrow $ $^{218}$Fr as
a function of the separation $r$ between the nuclei according to the TCSM. 
Levels are characterised by the total angular momentum projection $j_{z}$ on 
the 
internuclear axis: $j_{z}=1/2$ (dashed curves), $j_{z}=3/2$ (dotted curves), 
$j_{z}=5/2$ (dashed-dotted curves) and other values (solid curves). 
L$^{*}$ and H$^{*}$ denote the level of the last neutron of $^{9}$Be and 
the Fermi level of $^{209}$Bi, respectively. 
L denotes a level of $^{9}$Be occupied by two neutrons, while 
the other levels belong to $^{209}$Bi. 
We only consider transitions 
of the last neutron of $^{9}$Be at the avoided crossings 1 and 2. 
The A, B and C arrows 
indicate the touching configuration of the nuclei if no neck is formed, 
the position of the Coulomb barrier and the radius where 
a neck between the nuclei starts to form, respectively. 
See text and \protect\cite{Alexis} for further details.}
\end{figure}

\newpage

\begin{figure}
\begin{center}
\epsfig{file=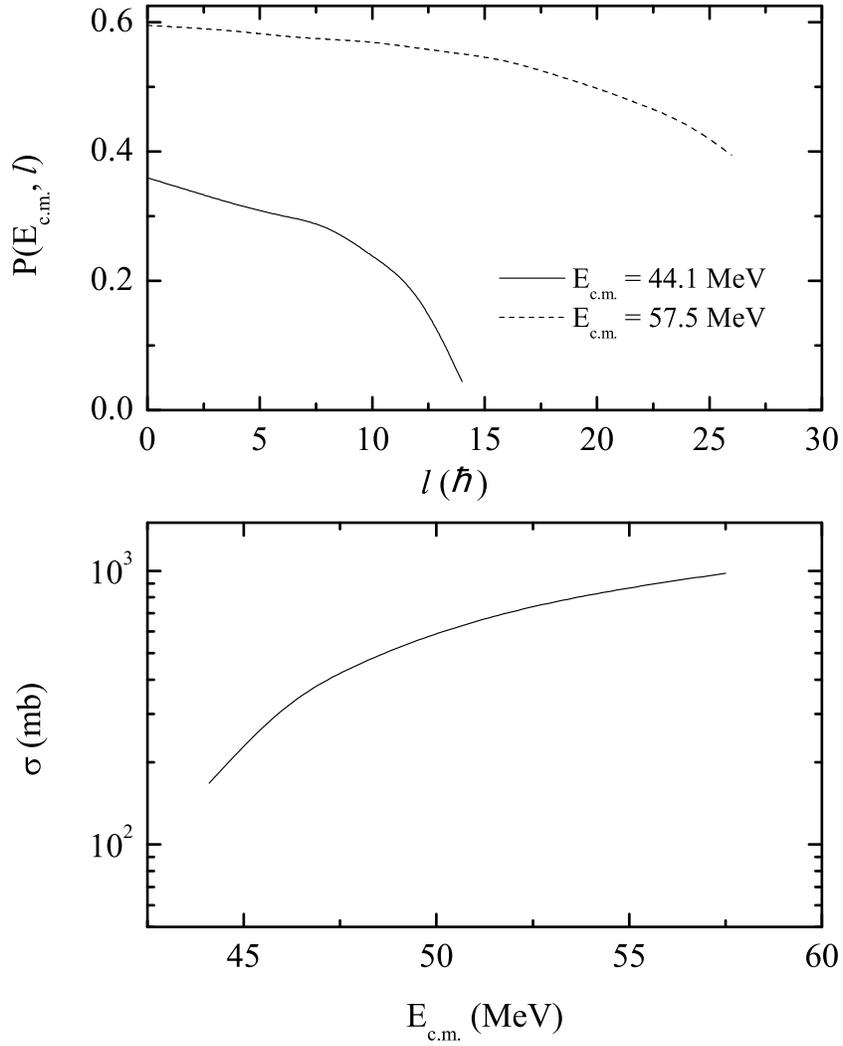,width=0.75\textwidth}
\end{center}
\caption{Dissociation probability $P(\text{E}_{\text{c.m.}},l)$ of $^{9}$Be 
on $^{209}$Bi as 
a function of the orbital angular 
momentum $l$ ($0\leq l\leq l_{max}$) (upper part). Values
for E$_{\text{c.m.}}=44.1$ MeV and E$_{\text{c.m.}}=57.5 $ MeV are shown 
by solid and dashed curves, respectively. See text for further details.
Calculated excitation function for the production 
of $\alpha$-particles for $^{9}$Be + $^{209}$Bi (lower part, solid curve).}
\end{figure}

\newpage

\begin{figure}
\begin{center}
\epsfig{file=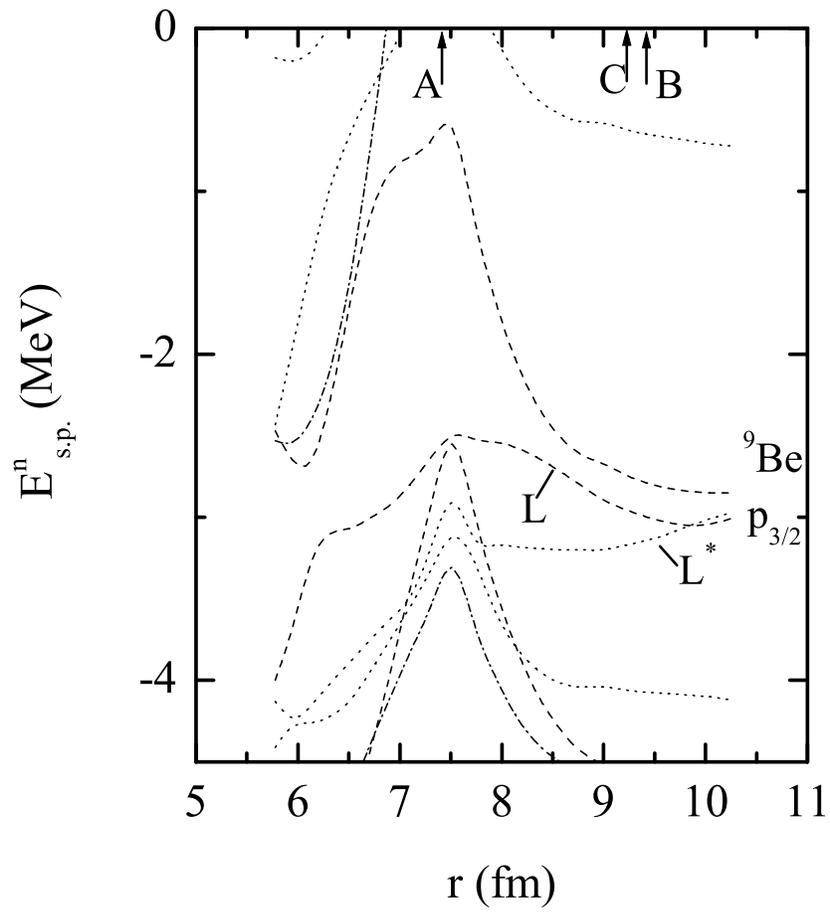,width=0.75\textwidth}
\end{center}
\caption{The same as in Fig. 1, but 
for $^{9}$Be + $^{64}$Zn $\rightarrow $ $^{73}$Se around the radius 
of the Coulomb barrier. See text for further details.}
\end{figure}

\newpage

\begin{figure}
\begin{center}
\epsfig{file=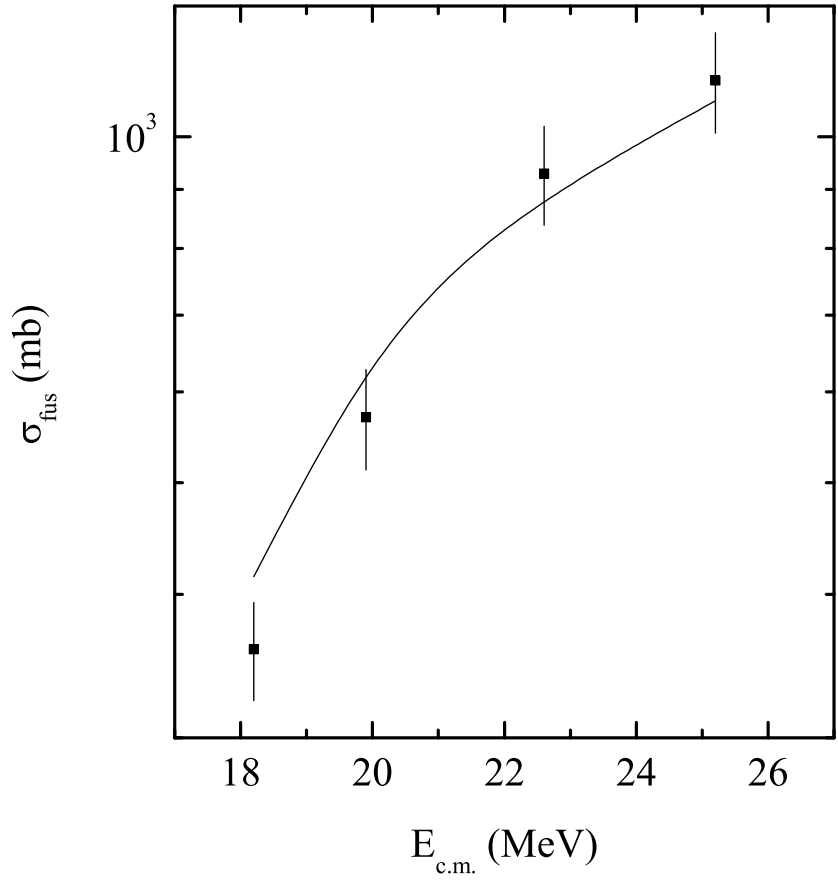,width=0.75\textwidth}
\end{center}
\caption{Complete fusion excitation function (solid curve) 
for $^{9}$Be + $^{64}$Zn calculated 
within the Glas-Mosel model and using the nucleus-nucleus potential obtained 
with the TCSM and the Strutinsky method. Experimental data (full squares) 
are from \protect\cite{Lubian}. See text for further details.}
\end{figure}

\end{document}